\documentclass[12pt,a4paper]{article}
\pdfoutput=1
\usepackage{cite}
\usepackage{amsmath}
\usepackage{color}
\usepackage{amsfonts}
\usepackage{amssymb}
\usepackage{graphicx}
\usepackage{geometry}
\usepackage{amssymb,epsfig,subfigure}
\usepackage{hyperref}
\usepackage{comment}
\usepackage[font=footnotesize]{caption}

\makeatletter
\renewcommand\section{\@startsection {section}{1}{\z@}%
                                 {-3.5ex \@plus -1ex \@minus -.2ex}
                                   {2.3ex \@plus.2ex}%
                                   {\normalfont\large\bfseries}}
\renewcommand\subsection{\@startsection{subsection}{2}{\z@}%
                                   {-3.25ex\@plus -1ex \@minus -.2ex}%
                                     {1.5ex \@plus .2ex}%
                                     {\normalfont\bfseries}}
\renewcommand\subsubsection{\@startsection{subsubsection}{3}{\z@}%
                                   {-3.25ex\@plus -1ex \@minus -.2ex}%
                                     {1.5ex \@plus .2ex}%
                                     {\normalfont\itshape}}
\makeatother

\def\pplogo{\vbox{\kern-\headheight\kern -29pt
\halign{##&##\hfil\cr&{\ppnumber}\cr\rule{0pt}{2.5ex}&\ppdate\cr}}}
\makeatletter
\def\ps@firstpage{\ps@empty \def\@oddhead{\hss\pplogo}%
  \let\@evenhead\@oddhead 
}
\def\maketitle{\par
 \begingroup
 \def\thefootnote{\fnsymbol{footnote}}
 \def\@makefnmark{\hbox{$^{\@thefnmark}$\hss}}
 \if@twocolumn
 \twocolumn[\@maketitle]
 \else \newpage
 \global\@topnum\z@ \@maketitle \fi\thispagestyle{firstpage}\@thanks
 \endgroup
 \setcounter{footnote}{0}
 \let\maketitle\relax
 \let\@maketitle\relax
 \gdef\@thanks{}\gdef\@author{}\gdef\@title{}\let\thanks\relax}
\makeatother

\numberwithin{equation}{section}

\newcommand\eea{\end{eqnarray}}
\newcommand\bea{\begin{eqnarray}}

\def\beq{\begin{equation}}
\def\eeq{\end{equation}}

\newcommand{\be}{\begin{equation}}
\newcommand{\ee}{\end{equation}}
\newcommand{\ba}{\begin{align}}
\newcommand{\ea}{\end{align}}
\newcommand{\bg}{\begin{gather}}
\newcommand{\eg}{\end{gather}}
\newcommand{\bseq}{\begin{subequations}}
\newcommand{\eseq}{\end{subequations}}

\renewcommand{\t}{\tilde}

\newcommand{\tr}{{\rm tr}}
\newcommand{\mc}{\mathcal}

\begin{document}
\setcounter{page}0
\def\ppnumber{\vbox{\baselineskip14pt
}}
\def\ppdate{
} \date{}

\author{Horacio Casini, Ignacio Salazar Landea, Gonzalo Torroba\\
[7mm] \\
{\normalsize \it Centro At\'omico Bariloche and CONICET}\\
{\normalsize \it S.C. de Bariloche, R\'io Negro, R8402AGP, Argentina}
}

\bigskip
\title{\bf  Irreversibility in quantum field theories \\ with boundaries
\vskip 0.5cm}
\maketitle

\begin{abstract}
We study conformal field theories with boundaries, and their boundary renormalization group (RG) flows, using methods from quantum information theory. Positivity of the relative entropy, together with unitarity and Lorentz invariance, give rise to bounds that characterize the irreversibility of such flows. This generalizes the recently proved entropic $g$-theorem to higher dimensions. In $2+1$ dimensions with a boundary, we prove the entropic $b$-theorem -- the decrease of the two-dimensional Weyl anomaly under boundary RG flows. In higher dimensions, the bound implies that the leading area coefficient of the entanglement entropy induced by the defect decreases along the flow. Our proof unifies these properties, and provides an information-theoretic interpretation in terms of the distinguishability between the short distance and long distance states.
Finally, we establish a sum rule for the change in the area term in theories with boundaries, which could have implications for models with localized gravity.

\end{abstract}
\bigskip

\newpage

\tableofcontents

\vskip 1cm

\section{Introduction}\label{sec:intro}

Boundaries are ubiquitous in broad areas of physics, including high energy, condensed matter physics and, of course, real experiments. In particular, systems that preserve scale invariance play a central role in the dynamics and phases of quantum field theories (QFTs) with boundaries. This is analogous to how fixed points of the renormalization group (RG) organize the dynamics of QFTs without boundaries \cite{Wilson:1973jj}. In this work we consider fixed points of relativistic QFTs with boundaries, and how they are connected by RG flows. Our goal is to understand if these flows are irreversible. We will approach the problem using quantum information theory.

The irreversibility of relativistic QFTs without boundaries has been established in two \cite{Zamolodchikov:1986gt, Casini:2004bw}, three \cite{Casini:2012ei}, and four \cite{Komargodski:2011vj, Casini:2017vbe} spacetime dimensions. Methods from quantum information theory, based on strong subadditivity of the entanglement entropy (EE), have provided a unifying proof for all these irreversibility theorems \cite{Casini:2017vbe}. A natural question is whether some of these properties survive in the presence of boundaries. 

The earliest result in this direction is the $g$-theorem: this establishes the irreversibility of boundary RG flows\footnote{A boundary RG flow is an RG flow triggered by a relevant deformation that is turned on only at the boundary. The bulk, as in all cases in this work, is conformal.} in 2d CFTs with boundaries \cite{Affleck:1991tk, Friedan:2003yc}. The entropic version of this result was recently obtained in \cite{Casini:2016fgb}. More generally, one can consider higher-dimensional quantum field theories with boundaries, that respect conformal invariance, and their boundary RG flows.\footnote{See \cite{Andrei:2018die} for a review with references.} In this paper we study generalizations of the approach of \cite{Casini:2016fgb} to higher-dimensional boundary conformal field theories (BCFTs).
We will use methods from quantum information theory to characterize boundary RG flows.

The introduction of boundaries poses new challenges to this approach. One is related to the (partial) breaking of the Poincar\'e group, which makes relativistic constraints less powerful. Another new issue is that, while the RG flow occurs at the boundary, the dynamics of the boundary degrees of freedom by themselves is nonlocal. The nonlocality is induced by interactions with the bulk, and it could prevent the existence of irreversibility theorems for such flows. A surprising outcome of our results will be that the nonlocality cancels out from the measures of quantum information theory that we analyze (the entanglement entropy and the relative entropy). While locality, causality and conformal invariance in the bulk are ultimately responsible for this, this is a point that remains to be fully understood and exploited.

Before proceeding, let us mention some encouraging results in higher dimensions. One is the $b$-theorem of \cite{Jensen:2015swa}, regarding the irreversibility of boundary RG flows in 3d. This was shown using dilaton methods \cite{Komargodski:2011vj}. There is also evidence for a boundary $F$-theorem in 4d systems with boundaries \cite{Gaiotto:2014gha}. On the other hand, holographic models also exhibit irreversibility of boundary RG flows in general $d$ \cite{Fujita:2011fp}. Other examples, and a suggestion for an irreversible quantity, are explored in the recent work \cite{Kobayashi:2018lil}.

This paper is organized as follows. In Sec.~\ref{sec:setup} we discuss relevant aspects of boundary RG flows, and how they can be measured in terms of the entanglement entropy. In Sec.\ref{sec:relE} we study the relative entropy and its connection with the entanglement entropy. We argue that, by taking the limit where the Cauchy surface becomes null, the modular hamiltonian contribution vanishes (in a certain range for the relevant deformation dimension). This is one of our main results, which allows us to use positivity of the relative entropy to bound the change in the EE between the UV and the IR. It extends \cite{Casini:2016fgb} to higher dimensions; its consequences are discussed in Sec.~\ref{sec:conseq}. We prove the entropic version of the $b$-theorem in 2+1 dimensions, and ``area theorems'' for higher dimensional QFTs with boundaries. In Sec.~\ref{sec:sum} we derive a sum rule that provides an explicit expression for the change in the EE in terms of the two-point function of the trace of the boundary stress tensor. Finally, Sec.~\ref{sec:concl} contains our conclusions and future directions.

\section{Boundary RG flows and entanglement entropy}\label{sec:setup}

We consider a quantum field theory (QFT) on a $d$-dimensional spacetime $\mathcal M\,:\,(x^0, \ldots x^{d-1})$ with a boundary $\partial \mathcal M$ at $y \equiv x^{d-1} =0$. The first $d-1$ coordinates $x^\alpha \in \mathbb R^{1,d-2}$, while $y$ lives on the half-line $y \ge 0$. The `bulk' fields have some prescribed boundary conditions at $y=0$, and there may also be degrees of freedom localized at the boundary.

We will be interested in boundary conformal field theories (BCFTs) and the RG flows that connect them.\footnote{See e.g. \cite{Billo:2016cpy} for a discussion of defects in CFTs.} The bulk is always taken to be a CFT, and the boundary breaks the bulk conformal group to
\be
SO(2, d)\, \to \,SO(2,d-1)\,.
\ee
In other words, a conformal boundary preserves the dilatation, Poincar\'e and special conformal transformations that keep the boundary $y=0$ fixed. We start from a UV boundary CFT $\mathcal T_0$, and trigger a boundary RG flow by turning on a relevant deformation on the boundary,
\be\label{eq:SRG}
S_{\mathcal T_1}= S_{\mathcal T_0}+ \int_{\partial \mathcal M}\,d^{d-1}x\, g\, \mathcal O
\ee
where the dimension of $\mathcal O$ at the UV fixed point is $\Delta_{\mc O} \le d-1$. The theory with relevant deformation is denoted by $\mathcal T_1$, and we assume that the flow ends at an IR fixed point -- a different BCFT.

We wish to understand if these flows are irreversible in some sense. In order to address this, we will employ methods from quantum information theory, generalizing the recent entropic proof of the $g$-theorem given in \cite{Casini:2016fgb}. The basic idea is to compute the entanglement entropy for a (half) sphere of radius $R$, centered at the boundary, and then use the monotonicity of the relative entropy in order to characterize the irreversibility of the RG. Before describing this in more detail, let us review the simpler case of the $g$-theorem.

For a 2d theory with boundary at $x^1=0$, the contribution to the thermal entropy due to the boundary can only decrease along boundary RG flows \cite{Affleck:1991tk, Friedan:2003yc}. This contribution, denoted in general as $\log g$, can also be captured by the entanglement entropy (EE) on an interval $x^1 \in (0, R)$,
\be
 S( \rho_R)= -{\rm tr}( \rho_R \log \rho_R)\,,
\ee
where $\rho_R$ is the vacuum density matrix $|0\rangle \langle 0|$ reduced to the interval. At a fixed point, this is of the form \cite{Calabrese:2009qy}
\be\label{eq:Sg}
 S(\rho_R) = \frac{c}{6} \,\log\frac{R}{\epsilon}+c_0+\log g\,,
\ee
where $c$ is the bulk CFT central charge, $\epsilon$ is a short distance cutoff, and $c_0$ is some regulator-dependent constant. While the regulator dependence does not allow to obtain $\log g$ directly from the EE, we can measure \textit{changes} in impurity entropy by subtracting the EE between the UV and IR fixed points, where the bulk contributions cancel out:
\be
 S(m R \ll 1) -S(m R \gg 1) = \log \frac{g_{UV}}{g_{IR}}\,.
\ee
Here $m$ is some characteristic mass scale for the RG flow, for instance, $m \sim g^{1/(d-1-\Delta_{\mc O})}$ in (\ref{eq:SRG}).
Ref. \cite{Casini:2016fgb} showed that positivity of the relative entropy implies $S(m R \ll 1) -S(m R \gg 1)\ge 0$. Furthermore, monotonicity of this quantity also allows to define a monotonic interpolating $g$-function. Our goal is to extend this entropic $g$-theorem to higher dimensions.

\begin{figure}[h!]
\begin{center}  
\includegraphics[width=0.4\textwidth]{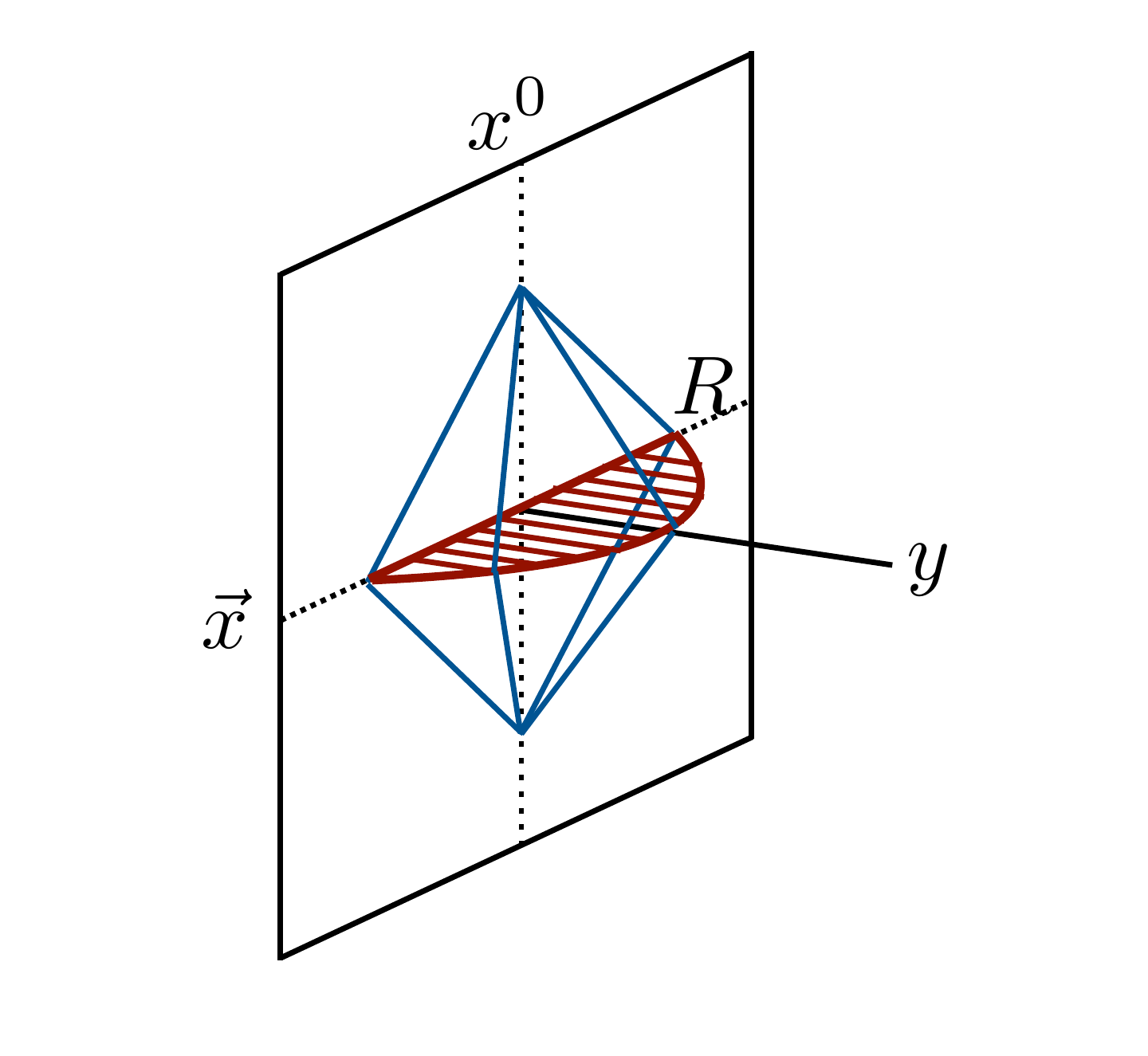}
\captionsetup{width=0.9\textwidth}
\caption{The entangling region is determined by a half-sphere of radius $R$ centered at the boundary $y=0$. The figure also shows the causal diamond of this region.
}
\label{fig:diamond}
\end{center}  
\end{figure}

Motivated by these results, we will consider the vacuum EE for (half) spheres
\be\label{eq:halfS}
(x^1)^2+ \ldots+(x^{d-2})^2+y^2 = R^2\;,\;y\ge 0\,,
\ee
in boundary field theories that undergo RG flows (\ref{eq:SRG}). This is shown in Fig. \ref{fig:diamond}, together with the associated causal domain. Unitarity and causality dictate that the EE depends on geometric properties of the boundary of the entangling region, and is the same for any Cauchy surface inside the causal diamond. As in \cite{Casini:2016fgb, Casini:2016udt}, our strategy will be to deform the Cauchy surface towards the light-cone in order to equate the EE with the relative entropy.

In more detail, let $\sigma$ be the vacuum reduced density matrix for the UV fixed point theory $\mc T_0$,
\be
\sigma_R= {\tr}_{\bar V}(|0\rangle \langle 0|)\,,
\ee
where $|0\rangle$ is the ground-state of $\mc T_0$, and $\bar V$ is the region outside the half-sphere (\ref{eq:halfS}). And let $\rho_R$ be the corresponding reduced density matrix for the theory $\mc T_1$ along the flow (\ref{eq:SRG}). The two theories are defined microscopically by the same operator content, but evolve with different hamiltonians. The relative entropy
\be
S(\rho_R|\sigma_R)={\tr_V}\left( \rho_R \,(\log \rho_R-\log \sigma_R)\right)
\ee
measures the distinguishability between the two states. It is positive and monotonic, namely it increases with the size of the region. Introducing the modular Hamiltonian of the UV BCFT,
\be
\sigma_R =\frac{e^{-\mc H_\sigma}}{{\rm tr}\,\mc H_\sigma}\,,
\ee
the relative entropy can be rewritten as
\bea\label{eq:Srel2}
S(\rho_R|\sigma_R)&=&{\rm tr}_V(\rho \mc H_\sigma)-{\rm tr}_V(\sigma \mc H_\sigma)+{\rm tr}_V(\rho_R \log \rho_R-\sigma_R \log \sigma_R ) \nonumber\\
&=& \Delta \langle \mc H_\sigma \rangle -\Delta S
\eea
with $\Delta \langle \mc H_\sigma \rangle={\rm tr}_V(\rho \mc H_\sigma)-{\rm tr}_V(\sigma \mc H_\sigma)$ and $\Delta S= S(\rho_R)- S(\sigma_R)$.

Following \cite{Casini:2016fgb}, the strategy will be to consider a Cauchy surface that approaches the light-cone of the causal diamond associated to the entangling region. We will show that, under certain circumstances, $\Delta \langle \mc H_\sigma \rangle \to 0$ in this limit. Then $S(\rho_R|\sigma_R) = -\Delta S$, and so the change in the EE inherits the positivity and monotonicity properties of the relative entropy. Before turning to this, though, let us discuss the general structure of the EE for BCFTs.

\subsection{Structure of the entanglement entropy}

In order to understand the structure of the EE for a half-sphere of radius $R$ in a BCFT, let us first recall that the sphere entropy in a CFT without boundary is of the form
\be\label{eq:EEsphere1}
S(R)=\mu_{d-2}\,R^{d-2}+\mu_{d-4}\, R^{d-4}+\ldots+ \left\lbrace \begin{array}{ll} (-)^{\frac{d-2}{2}} 4\,A\, \log(R/\epsilon)\,& d\;  \textrm{even}\,.\\ (-)^{\frac{d-1}{2}} F\,& d\,\,\textrm{odd} \,. \end{array}\right.
\ee
For a CFT the coefficients $\mu_{k}\sim \epsilon^{-k}$ are proportional to inverse powers of the cutoff. The geometric origin of these terms, for spheres or their null deformations, was recently explained in \cite{Casini:2018kzx}.  
The area term proportional to $R^{d-2}$ comes from a volume integral over the entangling surface, the term proportional to $R^{d-4}$ arises from a similar integral containing the intrinsic curvature of the surface, and so on. The nonlocal contribution, the logarithmic term in even $d$, comes from a Wess-Zumino action on the surface.

Moving on to the BCFT case, the entangling surface intersects the boundary  on a $d-3$ sphere of radius $R$. As in \cite{Casini:2018kzx}, this will give rise to new divergent terms in the EE, which can be written as integrals of local geometric quantities on this $S^{d-3}$ that respect Lorentz invariance. Therefore, for $d$ even (i.e. an odd $(d-1)$ - dimensional boundary), we have
\be\label{eq:Sbcfteven}
S(R) =\mu_{d-2}\,R^{d-2}+ \t \mu_{d-3}\,R^{d-3}+ \mu_{d-4}\, R^{d-4}+\ldots+(-1)^{\frac{d-2}{2}} 4\,A\, \log(R/\epsilon)+(-1)^{\frac{d-2}{2}} \t F\,,
\ee
where quantities that come from the boundary are denoted with a tilde. The last term behaves like a boundary F-function. Although the logarithmic term implies that $\t F$ will change when choosing different regulators, differences in $\t F$ for boundary RG flows will be physical. The simplest example is the case of the $g$-theorem discussed before, where $\log g$ in (\ref{eq:Sg}) plays the role of $\t F$. Examples of this for $d=4$ were studied in \cite{Gaiotto:2014gha}. The different coefficients are powers of the cutoff for a BCFT but can contain other divergent or finite contributions depending on the dimensionful scales of the theory as we approach a fixed point.  

For $d$ odd, the arguments in \cite{Casini:2018kzx} show that it is possible to have a Wess-Zumino action, now localized on the $S^{d-3}$, consistent with Lorentz invariance. Hence, we expect an entropy of the form
\be\label{eq:Sbcftodd}
S(R) =\mu_{d-2}\,R^{d-2}+ \t \mu_{d-3}\,R^{d-3}+ \mu_{d-4}\, R^{d-4}+\ldots+(-1)^{\frac{d-3}{2}} 4\,\t A\, \log(R/\epsilon)+(-1)^{\frac{d-1}{2}}  F\,.
\ee
The logarithmic term, absent for CFTs in odd dimensions, comes from the Wess-Zumino action on $S^{d-3}$. This should correspond to a Weyl anomaly in the BCFT. To see this, one can follow the arguments of \cite{Casini:2011kv} that related the logarithmic term in (\ref{eq:EEsphere1}) to the Weyl anomaly of the CFT. This involves a conformal mapping between the causal diamond of the spherical region and the static patch of de Sitter. The EE entropy on the sphere then becomes the thermal entropy of de Sitter. Rotating to euclidean time, this entropy is obtained from the logarithm of the partition function on $S^d$. The logarithmic term in the entropy is then seen to be determined directly by the Weyl anomaly
\be
\langle T^\mu_{\;\mu} \rangle =2(-1)^{d/2} A\,E_d +\ldots\,,
\ee
where $E_d$ is the Euler density,\footnote{The integral of $E_d$ on a $d$-dimensional sphere gives $2$.} and we are not showing terms that vanish for a sphere. Applying this to BCFTs with (\ref{eq:Sbcftodd}) maps the causal diamond of (\ref{eq:halfS}) to half of a causal patch in de Sitter. In the euclidean version, this is a hemisphere of $S^d$, with boundary at the equator. The logarithmic term in the thermal entropy should then come from a Weyl anomaly localized at the equator,
\be
\langle T^\mu_{\;\mu} \rangle =2(-1)^{\frac{d-1}{2}} \t A\,E_{d-1}\; \delta(\theta-\frac{\pi}{2}) +\ldots \,,
\ee
where $\theta \in (0, \pi/2)$ is the azimuthal angle of the hemisphere. See also \cite{Fursaev:2016inw, Kobayashi:2018lil, Herzog:2015ioa}.

The case of a 3d theory with boundary will play an important role below. There is a ``Graham-Witten'' anomaly from the 2d boundary \cite{Schwimmer:2008yh}
\be\label{eq:GW}
\langle T^\mu_{\;\mu} \rangle = - \frac{b}{24\pi}\,{\mc R}^{(2)}\,\,\delta(\theta-\frac{\pi}{2}) +\ldots
\ee
Then we have 
\be
\tilde{A}=\frac{b}{12}\,.\label{cero}
\ee
The $b$-anomaly decreases along boundary RG flows \cite{Jensen:2015swa}. Our goal will be to establish this, together with natural extensions to higher dimensions (area-theorems) using entropic methods.

\section{Analysis of the relative entropy}\label{sec:relE}

In this section we analyze the relative entropy (\ref{eq:Srel2}) between theories $\mc T_0$ and $\mc T_1$. After determining the structure of the modular Hamiltonian and properties of the stress tensor, we take the limit where the Cauchy surface approaches the light-cone of the causal domain.

\subsection{Modular Hamiltonian and stress-tensor}\label{subsec:modH}

For a CFT and a spherical region, the modular Hamiltonian is local, and given by
\be\label{eq:modH}
\mc H=\int_\Sigma\,d\sigma\,\eta^\mu\,\xi^\nu\,T_{\mu\nu}\,,
\ee
where $\eta^\mu$ is a unit vector normal to the Cauchy surface $\Sigma$, and $\xi^\nu$ is the Killing vector for a conformal transformation that keeps the sphere fixed,
\be\label{eq:killing}
\xi^\nu=\frac{\pi }{R} \left(R^2-(x^0)^2-(x^i)^2,-2 x^0 x^i \right)\,.
\ee
Recall that this can be proved by a conformal map between the causal domain of the sphere and the Rindler wedge \cite{Casini:2011kv,Cardy:2016fqc}; here the density matrix is thermal and the modular flow corresponds to boosts.

This approach can be applied to BCFTs as well to establish that (\ref{eq:modH}) is valid for a half-sphere centered at the boundary. The conformal transformation of \cite{Casini:2011kv} in $(-+\ldots +)$ signature reads 
\be\label{maa}
x^\mu=2\frac{X^\mu+(X\cdot X)C^\mu}{1+2(X\cdot C)+(X\cdot X)(C\cdot C)}-R^ 2C^\mu\,.
\ee 
We can perform it along the direction transverse to the boundary, $C^\mu \partial_\mu =R^{-1} \partial_y$. The boundary is then uniformly accelerated inside the Rindler wedge, with acceleration $1/R$. This respects the boost symmetry, and so the density matrix is again thermal with respect to the boost Hamiltonian. Therefore, the modular Hamiltonian is still given by the Rindler hamiltonian, and transforming back to the sphere gives
(\ref{eq:modH}).\footnote{See also \cite{Jensen:2013lxa} for this result; this work also discusses a map with $C_\mu$ tangential to the boundary.} We stress that $T_{\mu\nu}$ here is the stress-tensor operator of the BCFT $\mc T_0$, because we are using the conformal map between the sphere and the Rindler wedge.

Having established the form of the modular hamiltonian $\mc H_\sigma$ for the BCFT theory $\mc T_0$, let us analyze the contribution $\Delta \langle \mc H_\sigma \rangle$ to the relative entropy, for a given Cauchy surface $\Sigma$. Since $\xi^\nu T_{\mu\nu}$ is a conserved current, $\mc H_\sigma$ is a conserved charge on $\Sigma$. This means that, as an operator, $\mc H_\sigma$ is independent of the choice of Cauchy surface inside the given domain of dependence. All the dependence on $\Sigma$ has to come from the state. The key point, discussed in \cite{Casini:2016fgb, Casini:2016udt}, is that the state $\rho_R$ brings in explicit dependence on $\Sigma$. The reason is that, in order to compare both states $\rho$ and $\sigma$, we have to map the algebra of operators $\tilde \phi_\lambda(x)$ of $\mc T_1$ to that of $\mc T_0$, $\phi_\lambda(x)$, at a given Cauchy surface $\Sigma$. In the Heisenberg representation, these algebras evolve with different Hamiltonians, which means that the map depends on $\Sigma$,
\be
U_\Sigma\,\tilde \phi_\lambda(x) U_\Sigma^\dag = \phi_\lambda(x)\;,\;x \in \Sigma\,.
\ee
The corresponding map for the state, $\rho \to U_\Sigma \rho U_\Sigma^\dag$, will then also depend on the choice of the Cauchy surface.

To make this more explicit, let us consider the change in the stress-tensor expectation value between the two states,
\be
\Delta \langle T_{\mu\nu}(x) \rangle = {\rm tr}(\rho_R T_{\mu\nu}(x))-{\rm tr}(\sigma_R T_{\mu\nu}(x))\,.
\ee
Let us split the energy-momentum tensor into a bulk part, that is continuous at the boundary, and localized terms,
\be
T_{\mu\nu}(x) = T_{\mu\nu}^{bulk}(x)+ \delta(y)\,t_{\mu\nu}(x)\,,
\ee
with $t_{\mu\nu}$ associated to the presence of the boundary. First, away from the boundary $\rho_R$ and $\sigma_R$ evolve with the same Hamiltonian --they only differ by the relevant perturbation (\ref{eq:SRG}) localized at $y=0$. This means that $\Delta \langle T^{bulk}_{\mu\nu}(x) \rangle$ is independent on the choice of Cauchy surface. This is a local quantity, which can only depend on $y$ (there is translation invariance on the other coordinates), and on the metric $g_{\mu\nu}$ and the normal $N^\mu\partial_\mu=\partial_y$. We parametrize
\be\label{eq:bulk-ansatz}
\Delta \langle T_{\mu\nu}^{bulk}(x) \rangle= \alpha_1(y) (g_{\mu\nu}- N_\mu N_\nu)+ \alpha_2(y) N_\mu N_\nu\;.
\ee
Imposing the conservation condition $\partial_\mu T^\mu_y=0$ away from the defect sets $\partial_y \alpha_2=0$, so that $\alpha_2$ is constant. But since the expectation value has to decay away from the defect, we have $\alpha_2=0$. Also, $T_{\mu\nu}$ is the BCFT stress tensor and so it is traceless, $T^\mu_\mu=0$; this sets $\alpha_1=0$. Therefore, 
\be\label{eq:bulk-vanish}
\Delta \langle T^{bulk}_{\mu\nu}(x) \rangle=0\,,
\ee
and we are only left with a possible contribution proportional to $\delta(y)$. This is a key property of codimension one defects; as discussed in the Appendix, it is not valid for other codimensions.

On the boundary $\partial \mc M$, the state $\rho_R$ depends on the choice of Cauchy surface $\Sigma \cap \partial \mc M$. We can write the stress-tensor expectation value in terms of local geometric objects $g_{\mu\nu}, \, N_\mu, \, \eta_\mu$, and curvatures:
\be\label{eq:ansatz}
\Delta \langle T_{\mu\nu}(x) \rangle= \delta(y) \left(\alpha_1 (g_{\mu\nu}- N_\mu N_\nu)+ \alpha_2 \eta_\mu \eta_\nu+ \alpha_3 N_\mu N_\nu+ \alpha_4 (N_\mu \eta_\nu+N_\nu \eta_\mu) \right)+ O(\epsilon^2 K^2, \ldots)\,,
\ee
where we indicated that contributions from curvatures are suppressed by the short-distance cutoff $\epsilon$ \cite{Casini:2016udt}. In conformal perturbation theory around the UV fixed point we expect, on dimensional grounds,
\be\label{eq:scaling}
\alpha_i \sim g^2 \epsilon^{(d-1)-2\Delta_{\mc O}}
\ee
for $\Delta_{\mc O}\ge (d-1)/2$. For smaller operator dimension, a nonperturbative finite term of order $\alpha_i \sim g^{(d-1)/(d-1-\Delta_{\mc O})}$ can appear.

Plugging (\ref{eq:ansatz}) into (\ref{eq:modH}), and taking into account that $\xi^y=0$ at $y=0$, we find that the integral is restricted to the component $\Sigma \cap \partial \mc M$ of the entangling surface at the boundary, and that all surviving terms are proportional to the contraction $\eta^\alpha \xi_\alpha$:
\be
\Delta \langle \mc H_\sigma \rangle = \int_{\Sigma \cap \partial \mc M} d\sigma\,\alpha\,\eta^\alpha \xi_\alpha+O(\epsilon^2 K^2)\,.
\ee
Here $\alpha$ denotes a linear combination of the $\alpha_i$ in (\ref{eq:ansatz}); in what follows we will only need their order of magnitude (\ref{eq:scaling}). Note that $\Delta \langle \mc H_\sigma \rangle$ becomes an integral of a local quantity on the boundary.

We conclude that the dependence of $\Delta \langle \mc H_\sigma \rangle $ on $\Sigma$ is given simply by the flux of the conformal Killing vector $\xi^\mu$ through $\Sigma\cap \partial \mc M$. This is analogous to what happened for bulk RG flows in \cite{Casini:2016udt}, although here the bulk part is conformal, and all the contributions are restricted to the boundary.

\subsection{The light-cone limit}

The last step is to take a Cauchy surface that approaches the null boundary of the causal domain of dependence. The modular flow keeps the boundary of the causal domain fixed, so $\xi^\mu$ becomes null there. As a result,
\be\label{eq:null}
\xi^\alpha \eta_\alpha \Big |_{\Sigma_{null}} \to 0\,.
\ee
When this is valid, $\Delta \langle \mc H_\sigma \rangle \to 0$, and hence the change in the EE equals the relative entropy between the two states,
\be\label{eq:master}
S(\rho_R |\sigma_R)= - \Delta S\,.
\ee

However, the null limit has to be taken in a controlled way, because (\ref{eq:null}) multiplies a possibly divergent contribution (\ref{eq:scaling}). Following \cite{Casini:2016udt}, we can approach the null limit with a Cauchy surface forming a hyperboloid of radius $a$, obtaining
\be\label{eq:Hscaling1}
\Delta \langle \mc H_\sigma \rangle \,\sim\,(g^2 \epsilon^{d-1-2\Delta_{\mc O}})\,(a^2 R^{d-3})\,.
\ee
The first factor  here comes from the expansion (\ref{eq:scaling}), while $a^2 R^{d-3}$ is the contribution of $\int\,\xi^\alpha \eta_\alpha$ for a Cauchy surface that is a hyperboloid of curvature scale of order $a^{-1}$. This scaling with $R$ should be contrasted with the dependence $\Delta \langle \mc H_\sigma \rangle \sim R^{d-1}$ on a constant time slice, which always dominates over the area term $\Delta S \sim R^{d-3}$ associated to $\Sigma \cap \partial \mc M$. The null limit reduces the scaling with $R$ by two powers, replacing this by $a^2$.

Given a short distance cutoff $\epsilon$, we can have at most $a \sim \epsilon$, so (\ref{eq:Hscaling1}) becomes
\be\label{eq:Hscaling2}
\Delta \langle \mc H_\sigma \rangle \,\sim\,g^2\,R^{d-3}\, \epsilon^{d+1-2\Delta_{\mc O}}\,.
\ee
In this way, $\Delta \langle \mc H_\sigma \rangle \to 0$ in the null limit for relevant perturbations in the range
\be\label{eq:restriction}
\Delta_{\mc O} < \frac{d+1}{2}\,.
\ee
For a $(d-1)$-dimensional boundary, the perturbation is relevant if $\Delta_{\mc O} \le d-1$. For $d=2,3$, (\ref{eq:restriction}) then imposes no restrictions (it covers the whole range of relevant deformations), while for higher $d$ it does. In this regime, (\ref{eq:master}) applies. For $\Delta_{\mc O}>(d+1)/2$, the modular Hamiltonian contribution to the relative entropy will compete with the area term from the EE, and we cannot fix the sign of $\Delta S$.

\section{Consequences in different dimensions}\label{sec:conseq}

We now restrict to  (\ref{eq:restriction}), and analyze the consequences of (\ref{eq:master}).

First, in $d=2$ the EEs of the two states can be written as (see (\ref{eq:Sbcfteven}))
\bea
S(\sigma_R) &=& \frac{c}{6}\,\log \frac{R}{\epsilon}+c_0' + \log g_{UV} \nonumber\\
S(\rho_R) &=& \frac{c}{6}\,\log \frac{R}{\epsilon}+c_0' + \log g(R)\,,
\eea
where we have defined a running $g$-function $\log g(R)$. In the IR, $g(R) \to g_{IR}$, the impurity entropy for the IR BCFT. The relative entropy then reads
\be
S(\rho_R |\sigma_R)=- \log \,\frac{g(R)}{g_{UV}}\,.
\ee
Monotonicity of the relative entropy requires $g'(R)\le 0$. This is the entropic $g$-theorem proved in \cite{Casini:2016fgb}.

In $d=3$, corresponding to a 2d boundary, (\ref{eq:Sbcfteven}) gives the fixed point expression
\be
S(\sigma_R) = \mu_1\,\frac{R}{\epsilon} - F +\frac{ b_{UV}}{3} \,\log\,\frac{R}{\epsilon}\label{uno}
\ee
where we used $\t A_{UV} = b_{UV}/12$, the boundary central charge in (\ref{eq:GW}). The entropy $S(\rho_R)$ has a more complicated radial dependence along the RG flow, but near the IR fixed point,
\be
S(\rho_R) = \mu_1\,\frac{R}{\epsilon} - F+ \frac{b_{IR}}{3}\,\log(mR)- \frac{ b_{UV}}{3}\,\log(m \epsilon)\,.\label{dos}
\ee
Here $m \sim g^{1/(d-1-\Delta_{\mc O})}$ is a typical mass scale for the RG flow. Then the relative entropy for $m R \gg 1$ becomes
\be
S(\rho_R |\sigma_R) \approx  \frac{1}{3} (b_{UV}- b_{IR})\,\log(mR)\,.
\ee

Positivity of the relative entropy thus implies
\be\label{eq:bthm}
b_{UV} \ge b_{IR}\,,
\ee
so that the boundary $b$-anomaly decreases along boundary RG flow. The result (\ref{eq:bthm}) was proved using dilaton methods in \cite{Jensen:2015swa}.

Lastly, in higher dimensions and for long distances $m R \gg 1$, we have
\be
\Delta S = (\t \mu_{d-3}^{IR}-\t \mu_{d-3}^{UV})\,R^{d-3}+ \ldots
\ee
In the range (\ref{eq:restriction}), Eq. (\ref{eq:master}) together with positivity of the relative entropy imply that
\be\label{eq:areathm}
\t \mu_{d-3}^{UV} \ge \t \mu_{d-3}^{IR}\,.
\ee
This is a new result on the irreversibility of boundary RG flows, established with quantum information theory methods. It implies that the leading area term associated to the boundary can only decrease along boundary RG flows. From \cite{Casini:2014yca}, the flow in the area term in relativistic QFTs is related to the renormalization of Newton's constant. Thus, we expect that (\ref{eq:areathm}) may be relevant in theories of localized gravity \cite{Randall:1999vf}.

In summary, we have found that, in QFTs with boundaries, positivity of the relative entropy on a null Cauchy surface leads to an inequality that unifies the $g$-theorem, the $b$-theorem and an area theorem in higher dimensions. The change in these quantities thus acquires a precise information-theoretic meaning as a meassure of distinguishability between the reduced states of the UV and IR fixed points in boundary RG flows.

\section{A sum rule in boundary QFTs}\label{sec:sum}

For RG flows in relativistic QFTs without boundary, the change in the area term \be
\Delta \mu_{d-2}= \mu^{IR}_{d-2}-\mu^{UV}_{d-2}
\ee
between the UV and IR fixed points obeys a sum rule (see \cite{Rosenhaus:2014ula, Rosenhaus:2014zza, Casini:2014yca, Casini:2015ffa}),
\be
\Delta \mu_{d-2}=-\frac{\pi}{d(d-1)(d-2)}\,\int d^dx\,x^2\,\langle \Theta(x) \Theta(0) \rangle\,,
\ee
with $\Theta(x) = T^{\mu}_\mu(x)$. Given that for boundary RG flows we just established that 
\be
\Delta \tilde \mu_{d-3}=\t \mu_{d-3}^{IR}-\t \mu_{d-3}^{UV}  \le 0\,,
\ee
it is natural to ask whether a similar sum rule exists for this quantity.

In order to address this, let us follow \cite{Casini:2014yca}, and consider a half-sphere of radius $R \to \infty$. The entangling region approaches a Rindler wedge along (say) $x^1 \ge 0$. The boundary preserves boosts along $x^1$, so the modular Hamiltonian is given by
\be\label{eq:rindlerH}
\mc H =-2\pi\,\int_{w^1 \ge 0\;,\;w^{d-1} \ge 0}\, d^{d-1}\vec w\,w^1\,T_{00}(w)\,,
\ee
for any boundary QFT (not necessarily at a fixed point). Note that here $w^1=0$ is the Rindler edge, while $w^{d-1}=0$ is the position of the boundary.

Taking $R \to \infty$, $S(\rho_R)$ is dominated by the IR fixed point, and hence
\be
\Delta S = S(\rho_R)-S(\sigma_R) \approx (\t \mu_{d-3}^{IR}-\t \mu_{d-3}^{UV})\,R^{d-3}\,;
\ee
all the other terms are subleading. The change $\Delta \tilde \mu_{d-3}$ can be obtained by performing a small variation of $R$,
\be\label{eq:change1}
R\,\frac{d \Delta S}{dR}=(d-3)\,\Delta S\,.
\ee
Recall that under a small change of state $\delta \rho$, the first law allows to relate the variation in the entropy to the change in the modular hamiltonian, $\delta S =  {\rm tr}(\delta \rho\, \mc H )$. We can view (\ref{eq:change1}) as a small change in the state due to a dilatation. By applying the first law to the variation of entropy in each of the states we get
\bea\label{eq:DeltaS1}
\Delta S &=& \frac{1}{d-3}\,\int d^dx\,\langle \Theta_\rho(x) \mc H_{\rho} \rangle-\frac{1}{d-3}\,\int d^dx\,\langle \Theta_\sigma(x) \mc H_{\sigma} \rangle\nonumber\\
&=& \frac{1}{d-3}\,\int d^dx\,\delta(x^{d-1})\langle \theta_\rho(x) \mc H_{\rho} \rangle\,,
\eea
where we used that the trace of the stress-tensor vanishes in the BCFT state $\sigma$, and introduced
\be
\Theta_\rho(x) = \delta(x^{d-1}) \theta_\rho(x)\;,\;\theta_\rho ={\rm tr}( t_\rho)\,.
\ee 
Also, $\mc H_\rho$ and $\mc H_\sigma$ are the modular Hamiltonians of the theories $\mc T_1$ and $\mc T_0$ respectively.
Intuitively, $\int  d^dx\,\Theta(x)$ is implementing a global scale transformation, and its only nonvanishing contribution comes from the scale dependence at the boundary.

Replacing (\ref{eq:rindlerH}) into (\ref{eq:DeltaS1}), 
\be\label{eq:DeltaS2}
\Delta S=-\frac{2\pi}{d-3}\,\int d^dx\,\delta(x^{d-1})\,\int d^{d-1}w\,w^1\,\langle \theta(x) T_{00}(w) \rangle\,,
\ee
where the stress-tensors are evaluated in the theory $\mc T_1$, but we have eliminated the subindices $\rho$ to streamline the formulas. 

As we just reviewed, the factor of $\int d^dx\,\delta(x^{d-1}) \theta(x)$ is implementing a global scale transformation.
 The quantity $\int d^dx\,\delta(x^{d-1})\,\langle \theta(x) T_{\mu\nu}(w) \rangle$ is traceless and conserved in $w$ for bulk points, and therefore we can use the same arguments as in Sec.~\ref{subsec:modH} to conclude it has to vanish for $w$ in the bulk. In fact this is the change of the bulk stress tensor expectation value under dilatations, and the bulk stress tensor expectation value vanishes identically.  
 Only the boundary component of the stress-tensor will contribute to (\ref{eq:DeltaS2}), so here we can replace $T_{00}(w) \to \delta(w^{d-1}) t_{00}(w)$. In this way, we end up with a purely boundary expression
\be\label{eq:DeltaS3}
\Delta S=-\frac{2\pi}{d-3}\,\int d^dx\,\delta(x^{d-1})\,\int_{w^1\ge 0} d^{d-1}w\,\delta(w^{d-1})\,w^1\,\langle \theta(x) t_{00}(w) \rangle\,.
\ee
Translation invariance along the boundary implies that the integrand is independent of the spatial coordinates $(w^2,\,\ldots,\,w^{d-2})$ transverse to the Rindler edge $w^1=0$. These $w$-integrals give simply a factor of the boundary area $R^{d-3}$, and we arrive to
\be\label{eq:DeltaS4}
\Delta \t \mu_{d-3}=-\frac{2\pi}{d-3}\,\int_{\partial \mc M} d^{d-1}x\,\int_{w^1>0} dw^1\,w^1\,\langle \theta(x) t_{00}(w) \rangle\,.
\ee

Finally, we need to relate this to an integral of the two-point function of $\theta(x)$. Performing a diffeomorphism transformation tangent to the boundary (i.e. $\delta x^\mu(x^\alpha, x^{d-1})=v^\mu(x^\alpha, x^{d-1})$, with $v^{d-1}(x^\alpha, 0 )=0$) implies the conservation equation \cite{Friedan:2003yc}
\be
\partial_\alpha t^{\alpha \beta}=T_{bulk}^{\beta,d-1}\,.
\ee
We argued in Sec.~\ref{subsec:modH} that $T_{bulk}^{\mu\nu}=0$ in vacuum expectation value. Since in (\ref{eq:DeltaS4}) we have a global scale transformation on the vacuum expectation value $\langle t_{\alpha \beta}\rangle$, we can here use $\partial_\alpha t^{\alpha \beta}=0$ inside the correlation function.  In more detail,
\bea
\frac{\partial}{\partial w^\alpha}\int_{\partial \mc M} d^{d-1}x\,\langle \theta(x) t^{\alpha \beta}(w) \rangle =  R\frac{\partial}{\partial R} \langle   \frac{\partial}{\partial w^\alpha}t^{\alpha \beta}(w) \rangle=  R\frac{\partial}{\partial R} \langle   T_{bulk}^{\beta,d-1}(w) \rangle=0\,.
\eea
This  means that
\be
\int_{\partial \mc M} d^{d-1}x\langle \theta(x) t_{\alpha \beta}(w)\rangle=
\int_{\partial \mc M} d^{d-1}x (g_{\alpha \beta} \nabla^2-\partial_\alpha \partial_\beta) F(s)\,,
\ee
with $s=x-w$, and we used that this correlator is symmetric in $\alpha, \beta$. In particular, taking the trace over the $d-1$ coordinates of the boundary gives
\be
\langle \theta(x) \theta(w) \rangle = (d-2) \nabla^2 F(s)\,.
\ee

Let us now perform the integral,
\bea
\int_{\partial M} d^{d-1}x\,\int_{w^1>0}\,dw^1\,w^1\, \langle \theta(x) t_{00}(w) \rangle &=&\int_{\partial M} d^{d-1}x\,\int_{w^1>0}\,dw^1\,w^1\, \vec \nabla^2 F(s) \nonumber\\
&=& \int_{\partial M} d^{d-1}x\,F(x)\,,
\eea
where we integrated by parts twice. On the other hand,
\bea
\int_{\partial M} d^{d-1}x\,x^2\,\langle \theta(x) \theta(0)\rangle &=&(d-2)\int_{\partial M} d^{d-1}x\,x^2\, \nabla^2 F(x) \nonumber\\
&=&(d-1)(d-2) \int_{\partial M} d^{d-1}x\,F(x)\,,
\eea
after integration by parts.

Comparing both expressions, we derive the sum rule
\be
\Delta \t \mu_{d-3}=-\frac{2\pi}{(d-1)(d-2)(d-3)}\,\int_{\partial M} d^{d-1}x\,x^2\,\langle \theta(x) \theta(0)\rangle\,.
\ee
This sum rule agrees with the sign $\Delta \t \mu_{d-3} \le 0$ deduced from positivity of the relative entropy.

The case $d=3$ can be obtained by taking into account that in this case the behavior is logarithmic instead of a power law, or alternatively from the equation above by adimensionalizing the integral using a mass scale $m$
\be
\Delta S=-(m R)^{d-3}\,\frac{2\pi}{(d-1)(d-2)(d-3)}\,\int_{\partial M} d^{d-1}x\,m^{-(d-3)}x^2\,\langle \theta(x) \theta(0)\rangle\,,
\ee
and expanding for  $d\to3$ we have
\be
\Delta S=-\pi \log(m R) \,\int_{\partial M} d^2 x\,x^2\,\langle \theta(x) \theta(0)\rangle\,.
\ee
Comparing with (\ref{cero}), (\ref{uno}), and (\ref{dos}), this gives 
\be
b_{UV}-b_{IR}=3 \pi \,\int_{\partial M} d^2 x\,x^2\,\langle \theta(x) \theta(0)\rangle\,.
\ee
This coincides with the sum rule obtained in \cite{Jensen:2015swa} by a different method.

Note that this applies to $d \ge 3$, so that it does not provide a sum rule for the change in $g$ in $d=2$. The reason is that for $d=2$, the $\log( g_{IR }/g_{UV})$ contribution to $\Delta S$ does not scale like a power of $R$, and so it cannot be deduced from the Rindler limit. We can obtain the change in $\log g$ by integrating $\int dR\,(d \Delta S/dR)$, which makes it clear that the modular Hamiltonian over all scales $R$ would be needed.

\section{Conclusions and future directions}\label{sec:concl}

In this paper we proved a series of theorems about the monotonicity of the first subleading term in the entanglement entropy of CFTs with planar boundaries. We accomplished this by considering a null Cauchy surface, and equating the change in the EE with the relative entropy between the UV and IR states. This provides a unified picture for the $g$-theorem, $b$-theorem, and boundary area theorems in higher dimensions. Several questions arise on possible generalizations of our results. We will summarize our main future directions in what follows.

One direction is to consider defects with higher codimension. Our approach here does not apply to these cases because, as we review in the Appendix, the bulk contribution (\ref{eq:bulk-ansatz}) to the modular Hamiltonian does not vanish.
It is not clear yet how to resolve this issue. In this context, the recent work  \cite{Kobayashi:2018lil} conjectured that the defect free energy may provide a monotonic quantity for defect RG flows. For codimension one defects this agrees with the change $\Delta S$ in the EE entropy, but for higher codimensions both quantities are different. It would be interesting to find an information-theoretic version of this conjecture, and analyze its validity.

We saw that for $d=3$, the first subleading term in the EE of a BCFT corresponds to a logarithmic divergence. This is a universal term, in the sense that it is independent on the way we choose to regularize the theory, and has a nice interpretation in terms of the boundary Weyl anomaly $b$.
For $d>3$ the universal terms are even more subleading and we do not have access to their monotonicity properties using only the relative entropy. For CFTs without boundaries, one can access the universal terms in  $d=3, 4$ using the strong subaditivity of the EE and the Markovian property of the vacuum on the null cone. This gives the $F$-theorem and the entropic $a$-theorem \cite{Casini:2012ei, Casini:2017vbe}.
It would be interesting to explore if some of these results can be extended to BCFTs.

Finally, in Sec. \ref{sec:sum}  we have shown that the change of the subleading terms of the EE under an RG flow is related to a sum rule for the two point function of the trace of the boundary stress-energy tensor. In QFTs with full Poincare invariance,  Adler and Zee \cite{Adler:1982ri,Zee:1980sj} showed that this sum rule for the two point function is related to the renormalization of the Newton's constant. It would be interesting to understand if there are similar implications for the graviton effective action in field theories with boundaries.

\section*{Acknowledgments}
We thank Y. Sato for discussions.
This work was partially supported by CONICET (PIP grant 11220150100299), CNEA, and Universidad Nacional de Cuyo, Argentina. H.C. acknowledges an ``It From Qubit" grant of the Simons Foundation. G.T. is also supported by ANPCYT PICT grant 2015-1224.

\appendix

\section{Comments on higher-codimension defects}
\label{app:higher}

In this Appendix we briefly discuss some properties of the stress-tensor in theories with higher codimension defects. Most of this is standard and well-known; we include it to highlight how the approach in the main text fails when the codimension is not one.

We have a bulk CFT in $d$-dimensional flat spacetime, coupled to a defect of codimension $n$. We turn on a relevant deformation localized at the boundary; as before, $\sigma$ denotes the density matrix of the UV fixed point, while $\rho$ is the density matrix for the theory along the flow.

We split the coordinates as
\be
x_\mu=(x_\alpha, y_a)\;,\;\mu=0,\ldots, d-1\;;\;\alpha\;=0,\ldots, d-n-1\;;\; a=d-n,\ldots, d-1\,.
\ee
The defect is placed at
\be
y_a=0\,.
\ee
In the main part of this work we focused on $n=1$. Our goal is to evaluate $\Delta \langle T^{bulk}_{\mu\nu}(x) \rangle$, namely the change in expectation value of the defect CFT stress tensor between the states $\sigma$ and $\rho$. This will generalize the discussion in Sec.~\ref{subsec:modH} to other codimensions.

As discussed before, $\Delta \langle T^{bulk}_{\mu\nu}(x) \rangle$ is independent of the choice of Cauchy surface. Operatorially, $T^{bulk}_{\mu\nu}$ is conserved and traceless away from the defect. Furthermore, by rotational invariance it can only depend on $y= (y_a y_a)^{1/2}$. The conservation condition $
\partial_a T^{a\alpha}=0$ requires
\be
\Delta \langle T_{a \alpha} \rangle =0\,.
\ee 
The remaining nonzero components can then be parametrized as
\bea\label{eq:paramT}
\Delta \langle T_{\alpha \beta}\rangle&=& h(y) \eta_{\alpha \beta}\\
\Delta \langle T_{ab}\rangle&=& f_1(y) \delta_{ab}+f_2(y) \left(\frac{y_a y_b}{y^2}-\frac{\delta_{ab}}{n} \right)\,. \nonumber
\eea
Here we used Poincar\'e invariance along the defect to constrain $T_{\alpha \beta}$, and rotational invariance in the transverse directions to fix $T_{ab}$. 

Requiring vanishing trace relates
\be\label{eq:f1}
f_1(y)=-\frac{d-n}{n} h(y)\,.
\ee
The conservation condition $\partial_a T^{ab}=0$ gives, on the other hand,
\be\label{eq:f2}
\frac{n-1}{n} \left(f_2'(y)+\frac{n}{y} f_2(y) \right)+f_1'(y)=0\,.
\ee
So in general we have an arbitrary free function, which we may take as $h(y)$. 

The form of $h(y)$ is further determined if the vacuum state is conformally invariant: since there is no dimensionful coupling, and the stress tensor has dimension $d$, $h(y)= h_0/y^d$. As a result,
\bea\label{eq:T0}
\Delta \langle T_{\alpha \beta}\rangle &=&  \frac{h_0}{y^d}\,\eta_{\alpha \beta}\\
\Delta \langle T_{ab}\rangle &=& -\frac{d-n}{n}  \frac{h_0}{y^d}\,\delta_{ab}+\frac{d}{n-1}\frac{h_0}{y^d}\, \left(\frac{y_a y_b}{y^2}-\frac{\delta_{ab}}{n} \right)\nonumber\\
\Delta \langle T_{a \alpha} \rangle &=&0\,. \nonumber
\eea
Only the constant $h_0$ is arbitrary, and this depends on the type of conformal defect.

For $n=1$, the $f_2$ contribution in (\ref{eq:paramT}) vanishes identically, and then setting $f_2=0$ in (\ref{eq:f2}) requires $f_1$ to be a constant. This constant has to vanish because the one-point function decays away from the defect. Then $h(y)$ has to vanish as well, in order to satisfy (\ref{eq:f1}). This recovers $\Delta \langle T^{bulk}_{\mu\nu}(x) \rangle=0$ of the main text. On the other hand, we see that for higher codimensions $\Delta \langle T^{bulk}_{\mu\nu}(x) \rangle$ is non-vanishing. This means that the modular Hamiltonian contribution to the relative entropy will be generically nonzero and larger than $\Delta S$. The approach of Secs. \ref{sec:relE}, \ref{sec:conseq} does not apply here.

\bibliography{EE}{}
\bibliographystyle{utphys}

\end{document}